# Early formation of planetary building blocks inferred from Pb isotopic ages of chondrules

Jean Bollard[1], James N. Connelly[1], Martin J. Whitehouse[2], Emily A. Pringle[3], Lydie Bonal[4], Jes K. Jørgensen[1], Åke Nordlund[1], Frédéric Moynier[3] and Martin Bizzarro[1,3*]

[1]Centre for Star and Planet Formation, University of Copenhagen, Copenhagen, Denmark

[2]Department of Geosciences, Swedish Museum of Natural History, Stockholm, Sweden

[3]Institut de Physique du Globe de Paris, Université Paris Diderot, Sorbonne Paris Cité, France

[4]Institut de Planétologie et d'Astrophysique de Grenoble, Grenoble, France

*To whom correspondence should be addressed; bizzarro@snm.ku.dk.

ABSTRACT

The most abundant components of primitive meteorites (chondrites) are millimeter-sized glassy spherical chondrules formed by transient melting events in the solar protoplanetary disk. Using Pb-Pb dates of 22 individual chondrules, we show that primary production of chondrules in the early solar system was restricted to the first million years after formation of the Sun and that these existing chondrules were recycled for the remaining lifetime of the protoplanetary disk. This is consistent with a primary chondrule formation episode during the early high-mass accretion phase of the protoplanetary disk that transitions into a longer period of chondrule reworking. An abundance of chondrules at early times provides the precursor material required to drive the efficient and rapid formation of planetary objects via chondrule accretion.

INTRODUCTION

The discovery of thousands of exoplanets orbiting Sun-like stars (*1*) clearly establishes that planet formation is a ubiquitous process in the Galaxy. In the standard model (*2*), the formation of planets occurs in stages where small dust particles coalesce into 10-100 km diameter planetesimals, which collide to form planetary embryos and planets over timescales of 50 to 100 millions years (Myr). However, recent astronomical observations of young protoplanetary disks suggest much more rapid timescales for the growth of planetary cores. Indeed, detailed imaging of the <1 Myr HL Tau protoplanetary disk in the submillimeter/millimeter domain has revealed the presence of ring structures that are interpreted to reflect the early stages of planet formation (*3*). These rapid timescales are in keeping with newer models of planet formation where planetary growth is fuelled by pebble accretion, that is, the accretion of centimeter- to meter-sized particles loosely bound to the gas onto planetesimals seeds (*4*).

In the solar system, a record of the earliest stages of planet formation may be preserved in the most primitive meteorites (chondrites), which are fragments of asteroids that avoided melting and differentiation. The most abundant constituent of chondrites are chondrules, millimeter-sized glassy spherules formed as free-floating objects by transient heating events in the solar protoplanetary disk. Recent simulations indicate that the main growth of asteroids can result from the gas-drag-assisted accretion of chondrules (*5*), a process analogous to pebble accretion. In these models, the largest planetesimals of a population with a characteristic radius of ~100 km undergo run-away accretion of chondrules forming Mars-sized planetary embryos within a timescale of ~3 Myr. If chondrules represent the building blocks of planetary embryos and, by extension, terrestrial planets, understanding their chronology and formation mechanism(s) is critical to determine at which point during the early evolution of the solar system were conditions favourable to forming planetary bodies.

Of the various radiometric clocks, U-corrected Pb-Pb dating is the only method that provides a high-resolution assumption-free chronology of the first 10 Myr of the solar system. It is based on two isotopes of U that decay in a chain to stable Pb isotopes, namely $^{235}$U to $^{207}$Pb with a half-life of ~0.7 Gyr and $^{238}$U to $^{206}$Pb with a half-life of ~4 Gyr. Using this approach, it has been recently demonstrated that chondrule formation started contemporaneously with the condensation of the solar system's first solids – calcium-aluminum-rich inclusions (CAIs) – at 4567.3±0.16 Myr and lasted ~3 Myr (*6*). However, this chronological framework is based on only five individual objects such that it is not possible to provide a statistically meaningful analysis of the tempo and full duration of chondrule production. To provide an accurate chronology of chondrule formation based on a significant number of objects, we have determined the Pb-isotope compositions by thermal ionization mass spectrometry (TIMS) of 17 individual chondrules from primitive chondrite meteorites (Table 1), including the NWA 5697 ordinary chondrite and the NWA 6043 and 7655 CR2 carbonaceous chondrites. CR chondrites are considered to be one of the most primitive classes of meteorites, having experienced only mild aqueous alteration and showing no evidence for significant effects of thermal metamorphism (*7*). Moreover, analysis of the organic matter by Raman spectroscopy indicates that NWA 5697 is of petrologic type 3.10, that is, among the most pristine ordinary chondrites (see Supplementary Materials). The Pb isotope analyses enable us to derive Pb-Pb dates through the internal isochron approach by combining multiple fractions obtained by sequential acid dissolution of individual chondrules. These data are complemented by in situ analyses of the Pb isotope composition of the various mineral phases acquired by secondary ionization mass spectrometry (SIMS) for a subset of these chondrules, which allows us to determine the nature of the carrier phase of uranium and, hence, radiogenic Pb in these objects. Finally, using multiple collector plasma source mass spectrometry (MC-ICPMS), we have measured the $^{238}$U/$^{235}$U ratio of 7 individual chondrules (6 of which have been Pb-Pb dated) to test for potential U-isotope heterogeneity as well as the Zn stable isotope composition of 7 chondrules to assess the thermal history of their precursors.

**RESULTS**

**Pb-Pb isotopic ages of chondrules**

The subset of chondrules we investigated comprises all major petrographic classes, including both porphyritic and non-porphyritic texture types as well as FeO-rich and FeO-poor varieties (Table 1). The $^{238}$U/$^{235}$U ratio measured for seven individual chondrules from NWA 5697 returns values identical within analytical uncertainty to the solar $^{238}$U/$^{235}$U of 137.786±0.013 (Table 1, Fig. S24), confirming the lack of U-isotope variability amongst individual chondrules (*6,8,9*) at the resolution of our analyses. This observation validates the approach of using the solar $^{238}$U/$^{235}$U value to calculate Pb-Pb dates of individual chondrules, although for completeness we used the measured $^{238}$U/$^{235}$U values to calculate chondrule ages for which both U and Pb isotope compositions were determined. The six chondrules with measured $^{238}$U/$^{235}$U ratios define ages that span from 4567.61±0.54 Ma to 4564.65±0.46 Ma. This age range is comparable to that of 4567.57±0.56 Ma to 4563.24±0.62 Ma defined by the remaining 11 objects for which the solar $^{238}$U/$^{235}$U value is used to calculate the Pb-Pb ages. Using the bulk $^{238}$U/$^{235}$U of the host chondrites or, alternatively, bulk chondrule estimates of the same meteorites returns ages that are within 100,000 years of that obtained using the solar $^{238}$U/$^{235}$U value of 137.786±0.013 (Table S7) and, hence, well within the uncertainties of the final ages we report. To better understand the significance of our Pb-Pb dates, we have measured the Pb isotopic compositions of the individual mineral phases for chondrule 1-C2 (NWA 7655) and 3-C1 (NWA 5697) using in situ methods (see Supplementary Materials). Our analysis clearly establishes that the radiogenically produced Pb and, hence, U, is mainly present in the fine-grained chondrule mesostasis whereas the initial, non-radiogenic Pb is predominantly hosted by sulfides enclosed in the mesostasis (Fig. S25). That the radiogenic Pb is present in the mesostasis is in keeping with partition coefficient data as U typically behaves as an incompatible element in most silicates (*10*). These results are also in agreement with the systematics of the sequential acid dissolution approach used to define the Pb-Pb isochron ages, which reveals that the fractions with the most radiogenic Pb isotope compositions are released with the use of an acid mixture that preferentially dissolves non-refractory silicate phases (see Supplementary Materials). Thus, we infer that the Pb-Pb isochron ages of individual chondrules reflect the timing of crystallization associated with the last chondrule-melting event.

Our chondrule age dataset based on the analysis of 22 individual objects demonstrate that the production and melting of chondrules began contemporaneously with CAI condensation and melting and continued for ~4 Myr (Fig. 1A). We show in Fig. 1B the Pb-Pb age distribution of chondrules, which, at face value, indicates a progressive reduction in chondrule production rate through time. Indeed, approximately 50% of the chondrules investigated here formed within the first Myr of the protoplanetary disk, suggesting that chondrule formation was more efficient in early times. The residence time of millimeter-sized solids in evolving disks is predicted to be extremely short relative to the typical lifetimes of protoplanetary disks due to aerodynamic drag (*11*). Although some mechanisms exist to limit the inward drift of millimeter-sized solids such as, for example, dust trapping, these operate on timescales typically shorter than the age variability reported here for chondrules (*12*). Thus, the presence of an ancient component and significant age variability amongst chondrules from individual chondrites (Fig. 1) require effective outward mass transport and/or storage of chondrules during the lifetime of the protoplanetary disk. Recycling of early-formed refractory solids such as CAIs during chondrule-forming events has been observed in a number of primitive chondrites (*13-15*). Together with petrological evidence suggesting that many chondrules experienced multiple melting events (*16,17*), these observations raise the possibility that younger chondrule populations predominantly reflect remelting and, hence, recycling of chondrules formed at earlier times.

**Early chondrule formation and protracted recycling**

It is well-established that the flash heating events resulting in the melting of chondrules results in significant evaporative Pb loss and, hence, enhancement of the $^{238}U/^{204}Pb$ ratio (µ) relative to the solar value of ~0.19 given the refractory nature of U (*6*). This is consistent with the µ values for individual chondrules analyzed here that range from ~2 to ~183 (Table 1 and Supplementary Materials), which corresponds to up to ~99.94% of Pb loss relative to the solar composition. If the relatively high bulk µ values recorded by the majority of chondrules were acquired early, it would lead to the accumulation of substantial amounts of radiogenic Pb during the lifetime of the protoplanetary disk. Thus, chondrules with protracted complex thermal histories involving more than one melting episode are expected to record evolved initial Pb isotopic compositions relative to objects formed from precursors with near-solar µ values. Figure 2A shows the back projection of the regressions for each of the individual chondrules dated in this study, which allows us to assess the initial Pb isotope compositions of these objects. The chondrule dataset shows variable initial Pb isotope compositions corresponding to a range of ~120 ε-units in the $^{207}$Pb/$^{206}$Pb ratio, with the majority of chondrules recording evolved compositions relative to the most primitive composition defined by chondrules 2-C1 and C1. Incorporation of U-rich, refractory material such as CAIs in chondrule precursor could, in principle, produce apparently evolved initial Pb isotope compositions but this process would result in correlated variability between the initial Pb isotope compositions and the Al/Mg ratios of the bulk chondrules, which is not observed in our dataset (Fig. S26). A component of the elevated $^{207}$Pb/$^{206}$Pb ratio could also reflect mass-dependent heavy isotope enrichment associated with evaporative loss of Pb. To assess this possibility, we have measured the isotope composition of Zn, an element with comparable volatility as Pb (*18*), for seven chondrules. Assuming that the relative difference in the magnitude of the mass-dependent isotope fractionation between Zn and Pb is proportional to $1/m^2$ (*19*), Zn is predicted to record a level of isotopic fractionation that is ~10 times greater than Pb. The δ$^{66}$Zn values for the subset of chondrules range from −0.26±0.05 to −2.20±0.05‰, representing light isotope enrichment that is not expected during simple evaporative loss that typically results in heavy Zn isotope composition (*20*). The enrichment of light Zn isotopes in chondrules has been interpreted as reflecting removal of an isotopically heavy phase such as sulphide (*21*). This process is predicted to occur during remelting events (*22*), in keeping with the complex thermal reprocessing history inferred here. Regardless of the mechanism governing Zn isotope fractionation, the light and limited variability of Zn isotope fractionation in these chondrules (~2‰) confirms that the range of the $^{207}$Pb/$^{206}$Pb (~12‰) ratios is not a product of mass-dependent fractionation.

The initial Pb isotope composition of chondrules are positively correlated with their crystallization ages (Fig. 2B). Indeed, only chondrules formed in the first Myr of the protoplanetary disk record primitive compositions whereas younger chondrules show progressively more evolved initial Pb isotope compositions. The <1 Myr chondrule population records variability in the initial Pb isotope composition, which corresponds to ~50 ε-units in the $^{207}$Pb/$^{206}$Pb ratio. This range of initial Pb isotope compositions in the old chondrules could reflect the incorporation of radiogenic Pb possibly related to dust condensation following evaporative melting of CAIs, which can only occur at early times. The progressively more evolved initial Pb isotope compositions and the lack of primitive initial Pb recorded by younger chondrules suggest that these formed from precursors having already accumulated radiogenic Pb, which is indicative of material characterized by an elevated μ value. Apart from the chondrule-forming process, wide scale planetesimal melting and the production of magma oceans is an efficient mechanism for Pb devolatilization during the lifetime of the protoplanetary disk (*23*). Thus, it is conceivable that the evolved initial Pb isotope signatures observed in young chondrules reflect recycling of disrupted planetesimal fragments during chondrule forming episodes. However, the inferred µ values of differentiated planetesimals such as the angrite parent body are at least one order of magnitude more extreme than that observed in chondrules, inconsistent with the bulk of the precursor material of young chondrules reflecting disrupted planetesimal fragments. Moreover, the accretion, differentiation and establishment of long-lived magma oceans leading to planetesimal scale Pb devolatilization is thought to occur of over timescales of 3-4 Myr (*24,25*). Given that the majority of chondrules reported here have ages within ~3 Myr of solar system formation within analytical uncertainty, this process cannot easily account for the elevated µ values recorded by most chondrules. As such, we infer that the evolving initial Pb isotope compositions of chondrules during the lifetime of the disk reflects recycling and, hence, remelting of their precursors during earlier chondrule-forming events. Thermally-processed precursors may represent chondrule fragments or, alternatively, entire chondrules having experienced more than one chondrule-forming event.

A prediction of chondrule reworking in a closed system is a progressive increase in the inferred µ values of individual chondrules through time. Although such a correlation appears to be present in our dataset (Fig. S27), a number of different disk processes may perturb this relationship. For example, the chondrule-forming process may have been more efficient (with higher temperatures) at early times such that a progressive decrease in µ values with time would be expected for chondrules formed during the epoch of primary chondrule production, namely in the first Myr. Moreover, the accumulation of small amounts of Pb-rich dust onto the rims of chondrules that may be incorporated into the chondrule interior during flash melting will have the effect of lowering the µ value, which would prevent the µ from reaching extreme values in these cases. As such, it may be significant that the three chondrules projecting to modern terrestrial Pb isotopic compositions (and, therefore, having an effectively infinite µ value) correspond to older chondrules that may have avoided reworking and incorporation of Pb-rich dust.

To assess the validity of the proposal that the young, >1 Myr chondrules with evolved initial Pb isotope compositions reflect recycling of chondrules formed at early times, we back calculate the range of initial Pb isotope compositions that correspond to a U/Pb fractionation age of ~500,000 years after CAIs. This age is supported by the age distribution indicating that the bulk of the chondrules record crystallization ages within the first Myr of protoplanetary disk evolution (Fig. 1B). We show in Fig. 2B the range of initial Pb isotope compositions of the >1 Myr chondrules back calculated for a primary U/Pb fractionation age of 4566.8 Myr, which corresponds to ~50 ε-units in the $^{207}Pb/^{206}Pb$ ratio. This range of initial Pb isotope compositions is comparable to that of chondrules that record crystallization ages <1

Myr after solar system formation, consistent with the proposal that the bulk of the young chondrules reflect the recycling of chondrules formed at early times.

## DISCUSSION

**Absolute and relative chronology of chondrule formation**

The absolute ages we report demonstrate that the production and melting of chondrules began contemporaneously with CAI condensation and melting and continued for ~4 Myr. Apart from Pb-Pb dating, the only other method that has been utilized to provide insights into the timing of crystallization of individual chondrules via the internal isochron approach is the $^{26}$Al-to-$^{26}$Mg short-lived ($t_{1/2}$ ~ 0.7 Myr) decay system. Unlike the assumption free U-corrected Pb-Pb dating method, the validity of $^{26}$Al-$^{26}$Mg relative ages is based on the hypothesis that the $^{26}$Al/$^{27}$Al ratio was homogeneous in the solar protoplanetary disk with an initial value of ~5×10$^{-5}$ at the time of CAI formation. Attempts to provide a chronology of chondrule formation based on the short-lived $^{26}$Al-$^{26}$Mg system infers that chondrule formation began ~2-3 Myr after CAI condensation and melting (*26*). This apparent age difference, dubbed the CAI-chondrule age gap, has been used to constrain mechanisms of chondrule formation (*27*). We discuss below the possible explanations for the inconsistency between ages of individual chondrules derived by the Al-Mg and U-Pb systems.

One possible explanation for the age discordance is that the age variability or, alternatively, the preponderance of old ages inferred from Pb-Pb dating is an artifact of the progressive step-leaching dissolution technique utilized here relative to more traditional approaches based on the analyses of separated mineral phases. Three lines of evidence, however, suggest that this is not the case. First, two independent estimates of the Pb-Pb age for the SAH99555 angrite using the step-leaching dissolution technique (*28*) and a more traditional mineral separation approach (*29*) yield ages that are concordant within 280,000±405,000 years. Similarly, independent estimates for timing of condensation of CAIs using different techniques (*6,30*) define ages that are within 140,000±505,000 years. In both cases, the potential offset between the two techniques is well within the typical uncertainties of the chondrule ages reported here. Finally, using the step-leaching dissolution technique, Bollard *et al.* (*31*) report Pb-Pb ages for four individual chondrules from the Gujba metal-rich chondrite that are identical within 340,000 years. This chondrite is thought to have formed from a vapour-melt plume produced by a giant impact between planetary embryos, resulting in coeval ages of its various components (*32*). Collectively, these observations suggest that the Pb-Pb dates reported here for chondrules are accurate to within their stated uncertainties.

If the Pb-Pb dates are indeed accurate, a likely explanation for the age mismatch between the absolute and relative ages is that the assumption of homogenous initial disk $^{26}Al/^{27}Al$ composition that underpins the validity of $^{26}Al$-$^{26}Mg$ ages is incorrect. A number of recent studies have suggested a reduced initial $^{26}Al/^{27}Al$ value in solids that accreted to form protoplanets (*33,34*), which would translate into younger $^{26}Al$-$^{26}Mg$ ages relatively to the Pb-Pb dates. The bulk of the Pb-Pb ages reported here are from the NWA 5697 ordinary chondrite as well as various CR chondrites. We show in Fig. 3 the age distribution of NWA 5697 chondrules inferred from Pb-Pb dating relative to $^{26}Al$-$^{26}Mg$ ages of chondrules from the most primitive ordinary chondrites, assuming that the initial inner disk $^{26}Al/^{27}Al$ inventory was ~$1.5\times10^{-5}$ (*33*). Under this assumption, the two dating methods return comparable age distributions, consistent with the hypothesis of a reduced inner disk $^{26}Al$ inventory relative to the canonical abundance. However, the post 2 Myr age distribution inferred from the $^{26}Al$-$^{26}Mg$ system is difficult to characterise using existing data given the low $^{26}Al/^{27}Al$ abundance of $<2\times10^{-6}$ at that time. We note that preliminary results reporting the $^{26}Al$-$^{26}Mg$ and Pb-Pb ages of the same individual chondrules are consistent with the proposal of a reduced inner disk $^{26}Al$ inventory (*35,36*).

Finally, it has been recently suggested that metal-rich chondrites, including CR chondrules formed from an $^{26}Al$-poor reservoir possibly located beyond the orbits of the gas giant planets (*37,38*). In detail, it is proposed that the $^{26}Mg^*$ and $^{54}Cr$ compositions of CR chondrules require significant amounts (25-50%) of primordial $^{26}Al$-free molecular cloud matter in their precursor material. Accepting that the true age distribution of CR chondrules is reflected by their Pb-Pb systematics, a prediction of this model is that individual CR chondrules will record the lowest initial $^{26}Al/^{27}Al$ values relative to other chondrite groups and that a significant number of object will lack evidence for live $^{26}Al$ if these formed >1 Myr after CAI condensation. These predictions are in line with recent $^{26}Al$-$^{26}Mg$ systematics of CR chondrules, which record initial $^{26}Al/^{27}Al$ values of typically less than $3\times10^{-6}$ and more than 50% of the objects investigated having no sign for live $^{26}Al$ at the time of their crystallization (*39,40*). These observations, together with the clear evidence for $^{26}Al$ heterogeneity at the time of CAI formation (*41*), emphasize that the $^{26}Al$-$^{26}Mg$ system may not provide an accurate chronology of chondrule formation and, hence, disk processes.

**Dynamical evolution of the solar protoplanetary disk**

In contrast to non-igneous CAIs that formed as fine-grained condensates near the proto-Sun in a brief time interval associated with collapse of the presolar molecular cloud core, the majority of chondrules are thought to be products of flash heating of dust aggregates in different disk regions, possibly by shock wave heating (*42*). As shocks requires the existence of a gaseous disk, our chondrule Pb-Pb dates allow us to provide constraints on the lifetime of the solar protoplanetary disk, including the timing of its establishment relative to the proto-Sun, if indeed these objects are formed by shock-wave heating. Six chondrules (5-C1, 2-C1, 5-C2, 5-C10, 1-C1, C30) have ages identical to that of CAIs within analytical uncertainty, defining a population with a weighted mean age of 4567.41±0.19 Ma. Both astronomical observations and numerical simulations suggest that protoplanetary disks formed shortly after star formation, namely during the deeply-embedded stage (*43,44*). The age uncertainty associated with the ancient chondrule population indicate that these objects formed at most ~138,000 years after CAIs. Thus, this timescale defines an upper limit for the establishment of the protoplanetary disk formation after CAI formation and, by extension, the collapse of the proto-Sun. From an astronomical perspective, this time corresponds to the earliest deeply-embedded stage (class 0) of protostar evolution. It is well established that the disk was largely dissipated (*45*) at the time of formation of the impact-generated chondrules from the Gujba metal-rich chondrite at 4562.49±0.21 Ma (*31*). Using the age of the youngest chondrule identified here, we define minimum and maximum lifetimes of ~3.3 and ~4.5 Myr, respectively, for the active phase of the solar protoplanetary disk. These new timescales for the establishment and lifetime of the solar protoplanetary disk are in keeping with astronomical observations of young stars and theirs disks (*46*) and, importantly, indicates that the formation of a disk amiable to the production of asteroidal bodies and planetary embryos occurred shortly after collapse of the proto-Sun.

Our data and interpretation provide insights into the accretion history and thermal processing of dust in the protoplanetary disk. In the early stages of low mass star formation, mass accretion to the protostar occurs from the surrounding envelope via a circumstellar disk structure (*47*). This represents the deeply-embedded phase of star formation that only lasts for a small fraction of a disk lifetime, typically ~0.5 Myr compared to several million years. During this epoch, fresh, volatile-rich envelope material is processed through the disk and, thus, available to participate in the formation of early solar system solids. Chondrules formed within the first ~1 Myr of the disk lifetime have primitive initial Pb isotope compositions that are consistent with incorporation of thermally-unprocessed material characterised by a solar µ value. In contrast, chondrules formed at later times record evolved compositions, which require limited or no admixing of primordial dust with a solar µ value to their precursors. We infer that this reflects the transition between two distinct accretionary regimes during the early evolution of the solar protoplanetary disk. The first ~1 Myr reflects the main epoch of accretion and thermal processing of envelope material to the disk, which represents the regime of primary production of chondrules. As such, we suggest that the bulk of the chondrules preserved in chondrite meteorites were originally produced during this period. In contrast, the >1 Myr regime represents an epoch dominated by the transport and recycling of chondrules, including the remelting of first generation chondrules. The limited evidence for admixing of primordial dust with a solar µ value in >1 Myr chondrules indicates that the envelope of accreting material surrounding the proto-Sun had largely dissipated by that time.

**Chondrule formation and recycling mechanisms**

Based on petrographic, mineralogical and chemical observations, it is thought that chondrules formed by the melting of isotopically diverse solid precursors in dust-rich regions of the protoplanetary disk during repeatable and localised transient heating events (*27*). The inferred thermal histories of chondrules as well as the high solid densities (*48*) required for their formation are consistent with shock wave heating as the primary source of energy for the thermal processing and melting of chondrule precursors. Recent magnesium and chromium isotope systematics of CR chondrules suggest these objects formed in a reservoir distinct from other chondrite groups possibly located beyond the orbits of the gas giant planets (*37,38*). These data require a heat source enabling the production of chondrules at a wide range of orbital distances. A number of mechanisms have been proposed for producing chondrule-like objects in gaseous disks, including shocks produced by disk gravitational instability (*49,50*) as well as eccentrically orbiting planetesimals and planetary embryos (*51-53*). The Pb isotope systematics of individual chondrules reported here suggests that the primary chondrule production was limited to <1 Myr of the protoplanetary disk evolution, namely during the deeply-embedded stage of the proto-Sun charaterized by significant accretion of envelope material to the disk. The most efficient source of shocks during this epoch are shock fronts associated with spiral arms generated in a gravitationally unstable disk (*50, 54*). Numerical simulations indicate that this type of instability occurs in early formed disks that are relatively massive compared to their host stars ($M_d/M_* \geq 0.1$) (*55*). Shocks generated in this regime are modelled to be highly-efficient in the inner disk region and possibly extend to ~10 astronomical units (*50,56*), thereby providing a possible mechanism for the thermal processing of disk solids in early times on a global scale. However, as accretion to the disk decreases and the envelope dissipates, the resulting disk mass is thought to be far too low to sustain gravitational instabilities. Thus, a distinct source of shocks is required for remelting and, hence, recycling of the chondrules from ~1 to ~4 Myr after collapse of the proto-Sun. The accretion of large planetesimals and Mars-sized planetary embryos is believed to occur over timescales of ~0.5 to ~3 Myr (*24,33,57*), comparable to the lifetime of the protoplanetary disk inferred from our chondrule Pb-Pb dates. As such, bow shocks resulting from planetesimals and planetary embryos traveling on eccentric orbits provide a possible source of heating for the thermal processing of solids in the >1 Myr regime. Thus, it is apparent that different sources of heating are required to satisfy the Pb-Pb isotopic ages of individual chondrules, suggesting a multiplicity of chondrule-forming and remelting mechanisms. However, the higher proportion of <1 Myr chondrules with non-porphyritic textures (Fig. 1B), which indicates complete melting at higher temperatures, suggests that the chondrule-forming process was more efficient during early times.

The age variability recorded by individual chondrules establishes the existence of multiple generations of high-temperature solids within individual chondrite groups. This observation is in keeping with the existence of isotopic heterogeneity between chondrules from a single chondrite for nuclides such as Cr and Ti, which track genetic relationships between the silicate precursors of solids, asteroids and planetary bodies (*37,38,58*). Collectively, these data require that chondrule formation, recycling and outward mass transport occurred during the lifetime of the protoplanetary disk. Outward transport of chondrules could have occurred by a variety of time-dependent processes, including turbulent diffusion (*59*) and stellar outflows (*60*).

The observed chondrule age range does not support the concept that chondrules and matrix in a single chondrite group are genetically related, which is based on an apparent chemical complementarity between chondrules and coexisting matrix (*61*). Recent models of evolving viscous disks, however, suggest that a complementary relationship between chondrules and dust can be preserved for long time-scales provided that the decoupling between chondrules and gas is limited (*62*). In these models, various chondrule populations remained in complementarity such that the bulk contribution from each source is chemically solar and, thus, so is the final mixture. Alternatively, the observed chondrule-matrix complementarity may be an expression of the generic process of chondrule formation and does not reflect a genetic link (*38*). In this view, the matrix comprises a complement related to the chondrule formation process such that the bulk composition of the matrix is shifted from its starting composition and, thus, appears complementary to a chondrule composition. This does not require that the matrix is genetically linked to the chondrules in an individual chondrite but merely that some of it has experienced earlier chondrule formation events. Thus, fractions of the matrix in a particular chondrite may be complementary to chondrule populations in other chondritic meteorites. Likewise, the apparent isotopic complementarity for siderophile elements such as W and Mo (*63,64*) between chondrules and matrix may reflect the selective destruction and/or removal of isotopically anomalous metal phases during chondrule formation.

Finally, the efficient production of chondrule at early times and their continuous recycling inferred from our Pb isotope data is consistent with the proposal that chondrules may promote the growth of asteroidal bodies and planetary embryos by chondrule accretion (*5*). Chondrite meteorites are traditionally used to estimate the composition of the material that accreted to form the Earth. However, our results suggest that chondrules may be the dominant component controlling the composition of planetary bodies. Thus, a better understanding of the bulk elemental composition of these objects, including their volatile element inventory, may provide insights into the nature of the material precursor to terrestrial planets in our solar system and abroad.

**Materials and Methods**

Following identification of chondrules of suitable sizes from ~1 mm thick slabs of the NWA 5697, NWA 7655 and NWA 6043 meteorites, chondrules selected for isotopic investigations were characterized for their petrology and mineral chemistry using a scanning electron microscope and electron microprobe at the University of Copenhagen. Chondrules were liberated from the meteorite slabs using a variable speed Dremmel fitted with either cone-shaped, diamond-coated cutting tool or dental drill bits, and broken in fragments using an agate mortar and pestle. After sequential dissolution and chemical purification, the Pb isotope composition of each aliquot was determined using a ThermoFisher Triton TIMS at the Centre for Star and Planet Formation based on protocols described in (*6, 31*). We determined the Pb isotope composition of the individual phases of chondrules 1-C2 and 3-C1 using a CAMECA 1280 ion microprobe at the Swedish Museum of Natural History. The U isotope compositions of an aliquot of the chondrules used for Pb-Pb dating was determined for seven chondrules using the ThermoFisher Neptune Plus MC-ICPMS at the Centre for Star and Planet Formation based on protocols described in (*6*). Similarly, the Zn isotope compositions of an aliquot of the material used for Pb-Pb dating was determined for seven chondrules using the ThermoFisher Neptune Plus at the Institut de Physique du Globe de Paris based on protocols described in (*21*). All data needed to evaluate the conclusions in the paper are present in the paper and/or the Supplementary Materials.

**Supplementary Materials**

Fig. S12. Pb-Pb isochron diagram of Pb isotope analyses of NWA 5697 5-C4 chondrule.
Fig. S13. Pb-Pb isochron diagram of Pb isotope analyses of NWA 5697 3-C5 chondrule.
Fig. S14. Pb-Pb isochron diagram of Pb isotope analyses of NWA 5697 11-C1 chondrule.
Fig. S15. Pb-Pb isochron diagram of Pb isotope analyses of NWA 5697 11-C2 chondrule.
Fig. S16. Pb-Pb isochron diagram of Pb isotope analyses of NWA 5697 3-C2 chondrule.
Fig. S17. Pb-Pb isochron diagram of Pb isotope analyses of NWA 6043 1-C2 chondrule.
Fig. S18. Pb-Pb isochron diagram of Pb isotope analyses of NWA 6043 2-C2 chondrule.
Fig. S19. Pb-Pb isochron diagram of Pb isotope analyses of NWA 6043 2-C4 chondrule.
Fig. S20. Pb-Pb isochron diagram of Pb isotope analyses of NWA 6043 0-C1 chondrule.
Fig. S21. Pb-Pb isochron diagram of Pb isotope analyses of NWA 7655 1-C7 chondrule.
Fig. S22. Pb-Pb isochron diagram of Pb isotope analyses of NWA 7655 1-C2 chondrule.
Fig. S23. Pb-Pb isochron diagram of Pb isotope analyses of NWA 7655 1-C6 chondrule.
Fig. S24. $^{238}U/^{235}U$ ratios of individual chondrules, bulk chondrites, and achondrites.
Fig. S25. IMP data plotted in $^{204}Pb/^{206}Pb$ vs $^{207}Pb/^{206}Pb$ diagrams for chondrule NWA 7655 1-C2 (a) and chondrule NWA 5697 3-C1 (b).
Fig. S26. $^{27}Al/^{24}Mg$ and $\varepsilon(^{207}Pb/^{206}Pb)_{initial}$ variation diagram.
Fig. S27. Age and μ value variation diagram.
Fig. S28. Age and $\varepsilon(^{207}Pb/^{206}Pb)_{initial}$ variation diagram for selected chondrules.
Table S1. Electron microprobe analyses of the NWA 5697, NWA 7655, and NWA 6043 chondrules studied.
Table S2. Pb isotopic data for NWA 5697, NWA 7655, and NWA 6043 chondrules studied.
Table S3. Uranium isotopic compositions.
Table S4. Zinc isotopic compositions of selected chondrules.
Table S5. Ion-microprobe Pb analytical data for two chondrules.
Table S6. Summary of the chondrules investigated.
Table S7. Chondrule Pb-Pb ages corrected using different $^{238}U/^{235}U$ estimates.

**References and Notes**

1. F. Fressin, G.Torres, D. Charboneau, S.T. Bryson, J. Christiansen, C.D. Dressing, J.M. Jenkins, L.M. Walkowicz, N.M. Batalha, The false positive rate of Kepler and the occurrence of planets. *Astrophys J* **766**, 81 (2013).
2. J.E. Chambers, Planetary accretion in the inner Solar System. *Earth Planet Sci Lett* **223**, 241-252 (2004).
3. C. Carrasco-González, T. Henning, C.J. Chandler, H. Linz, L. Pérez, L. F. Rodríguez, R. Galván-Madrid, G. Anglada, T. Birnstiel, R. van Boekel, The VLA view of the HL Tau disk: Disk mass, grain evolution, and early planet formation. *Atrophy J Lett* **821**, L16 (2016).
4. B. Bitsch, M. Lambrechts, A. Johansen, The growth of planets by pebble accretion in evolving protoplanetary discs. *Astron Astrophys* **582**, A112 (2015).
5. A. Johansen, M.-M. Mac Low, P. Lacerda, M. Bizzarro, Growth of asteroids, planetary embryos, and Kuiper belt objects by chondrule accretion. *Science Adv* **1**, 1500109 (2015).
6. J.N. Connelly, M. Bizzarro, A. N. Krot, Å. Nordlund, D. Wielandt, M. A. Ivanova, The absolute chronology and thermal processing of solids in the solar protoplanetary disk. *Science* **338**, 651-655 (2012).
7. G. Briani, E. Quirico, M. Gounelle, M. Paulhiac-Pison, G. Montagnac, P. Beck, F.-R. Orthous-

Daunay, L. Bonal, E. Jacquet, A. Kearsley, S.S. Russell, Short duration thermal metamorphism in CR chondrites. *Geochim Cosmochim Acta* **122**, 267-279 (2013).

8. G.A. Brennecka, G. Budde, T. Kleine, Uranium isotopic composition and absolute ages of Allende chondrules. *Meteorit Planet Sci* **50**, 1995-2002 (2015).
9. J.N. Connelly, J. Bollard, M. Bizzarro, Pb-Pb chronometry and the early solar system. *Geochim Cosmochim Acta* **201**, 345-363 (2017).
10. J. Blundy, B. Wood, Mineral-melt partitioning of uranium, thorium and their daughters. *Rev Mineral Geochem* **52**, 59-123 (2003).
11. S.J. Weidenschilling, Aerodynamics of solid bodies in the solar nebula. *Mon Not Roy Astron Soc* **180**, 57-70 (1970).
12. L. Testi, T. Birnstiel, L. Ricci, S. Andrews, J. Blum, J. Carpenter, C. Dominik, A. Isella, A. Natta, J.P. Williams, D.J. Wilne, In *Protostars and Planets VI*, H. Beuther, R.S. Klessen, C.P. Dullemond, T. Henning, Eds. (Univ. Arizona Press, Tucson, 2014) pp. 339-361.
13. S. Itoh, H. Yurimoto, Contemporaneous formation of chondrules and refractory inclusions in the early Solar System. *Nature* **423**, 728-731 (2003).
14. A.N. Krot, H. Yurimoto, I.B. Hutcheon, G.J. MacPherson, Chronology of the early Solar System from chondrule-bearing calcium-aluminium-rich inclusions. *Nature* **434**, 998-1001 (2005).
15. A.N. Krot, K. Nagashima, E.M.M. van Kooten, M. Bizzarro, Calcium–aluminum-rich inclusions recycled during formation of porphyritic chondrules from CH carbonaceous chondrites. *Geochim Cosmochim Acta* **201**, 185-223 (2017).
16. A.N. Krot, J.T. Wasson, Igneous rims on FeO-rich and FeO-poor chondrules in ordinary chondrites. *Geochim Cosmochim Acta* **59**, 4951-4966 (1995).
17. A.N., Krot, G. Libourel, C.A. Goodrich, M.I. Petaev, Silica-rich igneous rims around magnesian chondrules in CR carbonaceous chondrites: Evidence for fractional condensation during chondrule formation. *Meteorit Planet Sci* **39**, 1931-1955 (2004).
18. K. Lodders, Solar system abundances and condensation temperatures of the elements. *Astrophys J* **591**, 1220-1247, (2003).
19. E.A. Schauble, Applying stable isotope fractionation theory to new systems. *Rev Mineral Geochem* **55**, 61-115 (2004).
20. F. Moynier, P. Beck, F. Jourdan, Q.-Z. Yin, Uwe Reimold, C. Koeberl, Isotopic fractionation of zinc in tektites. *Earth Planet Sci Lett* **277**, 482-489 (2009).
21. E. Pringle, F. Moynier, P. Beck, R. Paniello, D. Hezel, The origin of volatile element depletion in early solar system material: Clues from Zn isotopes in chondrules. *Earth Planet Sci Lett* **468**, 62-71 (2017).
22. J.T. Wasson, A.E. Rubin, Metal in CR chondrites. *Geochim Cosmochim Acta* **74**, 2212-2230 (2010)
23. J.N. Connelly, M. Bizzarro, Lead isotope evidence for a young formation age of the Earth-Moon system. *Earth Planet Sci Lett* **452**, 36-43 (2016).
24. M. Schiller, J. Baker, J. Creech, C. Paton, M.-A. Millet, A. Irving, M. Bizzarro, Rapid timescales for magma ocean crystallisation on the howardaite-eucrite-diogenite parent body. *Astrophys J* **740**, L22 (2011).
25. Y. Amelin, U-Pb ages of angrites. *Geochim Cosmochim Acta* **72**, 221-232 (2008).
26. A.N. Krot, Y. Amelin, P. Bland, F.J. Ciesla, J. Connelly, A.M. Davis, G.R. Huss, I.D. Hutcheon, K. Makide, K. Nagashima, L.E. Nyquist, S.S. Russell, E.R.D. Scott, K. Thrane, H. Yurimoto, Q.-Z. Yin, Origin and chronology of chondritic components: A review. *Geochim Cosmochim Acta*
Page **14** of 23

**73**, 4963-4997 (2009).

27. E.R.D. Scott, Chondrites and the protoplanetary disk. *Annu Rev Earth Planet Sci* **35**, 577-620 (2007).

28. J.N. Connelly, M. Bizzarro, K. Thrane, J.A. Baker, The Pb–Pb age of Angrite SAH99555 revisited. *Geochim Cosmochim Acta* **72**, 4813-4824 (2008).

29. Y. Amelin, The U–Pb systematics of angrite Sahara 99555. *Geochim Cosmochim Acta* **72**, 4874-4885 (2008).

30. Y. Amelin, A. Kaltenbach, T. Iizuka, C.H. Stirling, T.R. Ireland, M. Petaev, S.B. Jacobsen, U–Pb chronology of the Solar System's oldest solids with variable $^{238}$U/$^{235}$U. *Earth Planet Sci Lett* **300**, 343-350 (2010).

31. J. Bollard, J.N. Connelly, M. Bizzarro, Pb-Pb dating of individual chondrules from the CBa chondrite Gujba: Assessment of the impact plume formation model *Meteorit Planet Sci* **50**, 1197-1216 (2015).

32. A. N. Krot, Y. Amelin, P. Cassen, A. Meibom, Young chondrules in CB chondrites from a giant impact in the early Solar System. *Nature* **436**, 989-992 (2005).

33. M. Schiller, J.N. Connelly, A.C. Glad, T. Mikouchi, M. Bizzarro, Early accretion of protoplanets inferred from a reduced inner solar system $^{26}$Al inventory. *Earth Planet Sci Lett* **420**, 44-54 (2015).

34. K.K. Larsen, A. Trinquier, C. Paton, M. Schiller, D. Wielandt, M.A. Ivanova, J.N. Connelly, Å. Nordlund, A.N. Krot, M. Bizzarro, Evidence for magnesium isotope heterogeneity in the solar protoplanetary disk. *Astrophys J Lett* **735**, L37 (2011).

35. M. Bizzarro, M. Olsen, S. Itoh, N. Kawasaki, M. Schiller, L. Bonal, H. Yurimoto, Evidence for a reduced initial abundance of $^{26}$Al in chondrule forming regions and implications for the accretion timescales of protoplanets. *Meteorit Planet Sci* **49** (Suppl.), abstract #A43 (2014).

36. J. Bollard, N. Kawaski, N. Sakamoto, K. Larsen, D. Wielandt, M. Schiller, J. Connelly, H. Yurimoto, M. Bizzarro, Early disk dynamics inferred from isotope systematics of individual chondrules. *Meteorit Planet Sci* **50** (Suppl.) abstract #5211 (2015).

37. E.M.M.E. Van Kooten, D. Wielandt, M. Schiller, K. Nagashima, A. Thomen, K.K. Larsen, M.B. Olsen, Å. Nordlund, A.N. Krot, M. Bizzarro, Isotopic evidence for primordial molecular cloud material in metal-rich carbonaceous chondrites. *Proc. Natl. Acad. Sci. USA* **113**, 2011-2016 (2016).

38. M.B. Olsen, D. Wielandt, M. Schiller, E.M.M.E. Van Kooten, M. Bizzarro, Magnesium and $^{54}$Cr isotope compositions of carbonaceous chondrite chondrules - Insights into early disk processes. *Geochim Cosmochim Acta* **191**, 118 (2016).

39. D.L. Schrader, K. Nagashima, A.N. Krot, R.C. Ogliore, Q.-Z. Yin, Y. Amelin, C.H. Stirling, A. Kaltenbach, Distribution of $^{26}$Al in the CR chondrite chondrule-forming region of the protoplanetary disk. *Geochim Cosmochim Acta* **201**, 275-302 (2017).

40. K. Nagashima, A.N. Krot, G.R. Huss, $^{26}$Al in chondrules from CR2 chondrites. *Geochem J* **48**, 561-570 (2014).

41. J.C. Holst, M.B. Olsen, C. Paton, K. Nagashima, M. Schiller, D. Wielandt, K.K. Larsen, J.N. Connelly, J.K. Jørgensen, A.N. Krot, Å. Nordlund, M. Bizzarro, $^{182}$Hf-$^{182}$W age dating of a $^{26}$Al-poor inclusion and implications for the origin of short-lived radioisotopes in the early Solar System. *Proc Natl Acad Sci* **110**, 8819-8823 (2013)

42. F.J. Ciesla, L.L. Hood, The nebular shock wave model for chondrule formation: Shock processing in a particle-gas suspension. *Icarus* **158**, 281-293 (2002).

43. K. Tomida, S. Okuzumi, N.M. Machida, Radiation magnetohydrodynamic simulations of

protostellar collapse: non-ideal magnetohydrodynamic effects and early formation of circumstellar disks. *Astrophys J* **801**, 117 (2015).

44. J.J. Tobin, L.W. Looney, D.J. Wilner, W. Kwon, C.J. Chandler, T.L. Bourke, L.Loinard, H-F. Chiang, S. Schnee, X. Chen, A Sub-arcsecond survey toward Class 0 protostars in Perseus: Searching for signatures of protostellar disks. *Astrophys J* **805**, 125, (2015).

45. M.A. Morris, L.A. Garvie, L.P. Knauth, New insight into the solar system's transition disk phase provided by the metal-rich carbonaceous chondrite Isheyevo. *Astrophys J Lett* **801**, L22 (2015).

46. N.J. Evans, M.M. Dunham, J.K. Jørgensen, M.L. Enoch, B. Merín, E.F. van Dishoeck, J.M. Alcalá, P.C. Myers, K.R. Stapelfeldt, T.L. Huard, L.E. Allen, P.M. Harvey, T. van Kempen, G.A. Blake, D.W. Koerner, L.G. Mundy, D.L. Padgett, A.I. Sargent, The Spitzer c2d legacy results: Star-formation rates and efficiencies; evolution and lifetimes. *Astrophy J Supp* **181**, 321-350 (2009).

47. J.P. Williams, L.A. Cieza, Protoplanetary Disks and Their Evolution. *Annu Rev Earth Planet Sci* **49**, 67-117 (2011).

48. C.M.O'D. Alexander, J.N. Grossman, D.S. Ebel, F.J. Ciesla, The formation conditions of chondrules and chondrites. *Science* **320**, 1617-1619 (2008).

49. A.P. Boss, R.H. Durisen, Chondrule-forming shock fronts in the solar nebula: A possible unified scenario for planet and chondrite formation. *Astrophys J* **621**, L137-L140 (2005).

50. A.C. Boley, R.H. Durisen, Gravitational instabilities, chondrule formation, and the FU orionis phenomenon. *Astrophys J* **685**, 1193-1209 (2008).

51. L.L. Hood, M.A. Morris, N.A. Artemieva, F. Marzar, S.J. Weidenschilling, Nebular shock waves generated by planetesimals passing through Jovian resonances: Possible sites for chondrule formation. *Meteorit Planet Sci* **44**, 327-342 (2009).

52. M.A. Morris, F.J. Ciesla, S.J. Desch, T. Athanassiadou, Chondrule formation in bow shocks around eccentric planetary embryos. *Astrophys J* **752**, 27 (2012).

53. A.C. Boley, M.A. Morris, S.J. Desch, High-temperature processing of solids through solar nebular bow shocks: 3D radiation hydrodynamics simulations with particles. *Astrophys J* **776**, 101 (2013).

54. L. M. Pérez, J.M. Carpenter, S.M. Andrews, L. Ricci, A. Isella, H. Linz, A.I. Sargent, D.J. Wilner, T. Henning, A.T. Deller, C.J. Chandler, C.P. Dullemond, J. Lazio, K.M. Menten, S.A. Corder, S. Storm, L.Testi, M. Tazzari, W Kwon, N. Calvet, J.S. Greaves, R.J. Harris, L.G. Mundy, Spiral density waves in a young protoplanetary disk. *Science* **353**, 1519-1521 (2016).

55. K. Kratter, G. Lodato, Gravitational instabilities in circumstellar disks. *Annu Rev Earth Planet Sci* **54**, 271-311 (2016).

56. S.J. Desch, M.A. Morris, H.C. Connolly, A.P. Boss, The importance of experiments: Constraints on chondrule formation models. *Meteorit Planet Sci* **47**, 1139-1156 (2012).

57. N. Dauphas, A. Pourmand, Hf-W-Th evidence for rapid growth of Mars and its status as a planetary embryo. *Nature* **473**, 489-492 (2011).

58. A. Trinquier, T. Elliott, D. Ulfbeck, C. Coath, A. N. Krot, M. Bizzarro, Origin of nucleosynthetic isotope heterogeneity in the solar protoplanetary disk. *Science* **324**, 374-376 (2009).

59. F. Ciesla, Outward transport of high-temperature materials around the midplane of the solar nebula. *Science* **318**, 613-615 (2007).

60. S.H. Shu, H. Shan, T. Lee, Towards an astrophysical theory of chondrites. *Science* **271**, 1545-1552 (1996).

61. D.C. Hezel, H. Palme, The chemical relationship between chondrules and matrix and the chondrule-matrix complementarity. *Earth Planet Sci Lett* **294**, 85-93 (2010).

62. A.Z. Goldberg, J.E. Owen, E. Jacquet, Chondrule transport in protoplanetary discs. *Mon Not R Astron Soc* **452** 4054-4069 (2015).
63. G. Budde, T. Kleine, T.S. Kruijer, C. Burkhardt, K. Metzler, Tungsten isotopic constraints on the age and origin of chondrules. *Proc. Natl. Acad. Sci. USA* **113**, 2886-2891 (2016).
64. G. Budde, C. Burkhardt, G.A. Brennecka, M. Fischer-Gödde, T.S. Kruijer, T. Kleine, Molybdenum isotopic evidence for the origin of chondrules and a distinct genetic heritage of carbonaceous and non-carbonaceous meteorites. *Earth Planet Sci Lett* **454**, 293-303 (2016).
65. G.R. Huss, G.J. MacPherson, G.J. Waserburg, S.S. Russell, G. Srinivasan, Aluminum-26 in calcium-aluminum-rich in inclusions and chondrules from unequilibrated ordinary chondrites. *Meteorit Planet Sci* **36**, 975-997 (2001).
66. S. Mostefaoui, N.T. Kita, S. Togashi, S. Tachibana, H. Nagahara, Y. Morishita, The relative formation ages of ferromagnesian chondrules inferred from their initial aluminum-26/aluminum-27 ratios. *Meteorit Planet Sci* **37**, 421-438 (2002).
67. N.G. Rudraswami, J.N. Goswami, $^{26}$Al in chondrules from unequilibrated L chondrites: Onset and duration of chondrule formation in the early solar system. *Earth Planet Sci Lett* **257**, 231-244 (2007).
68. Villeneuve, J., M. Chaussidon, G. Libourel, Homogeneous distribution of $^{26}$Al in the solar system from the Mg isotopic composition of chondrules. *Science* **325**, 985-988 (1996).
69. J.L. Pouchou, F. Pichoir, Les elements tres legers en microanalyse X, possibilités des modèles récentes de quantification. *Microsc Spectrosc Electron* **11**, 229-225 (1986).
70. J.N. Connelly, M. Bizzarro, Pb-Pb dating of chondrules from CV chondrites by progressive dissolution. *Chem Geol* **259**, 143-151 (2009).
71. Y. Amelin, A. Ghosh, E. Rotenberg, Unravelling evolution of chondrite parent asteroids by precise U-Pb dating and thermal modeling. *Geochim Cosmochim Acta* **69**, 505-528 (2005).
72. H. Gerstenberger, G. Haase, A highly effective emitter substance for mass spectrometric Pb isotope ratio determinations. *Chem Geology* **136**, 309-312 (1997).
73. K.R. Ludwig, PBDAT: A computer program for processing Pb–U–Th isotope data. *United States Geological Survey Open File Report*. 88–524 (1993).
74. K.R. Ludwig, Isoplot/Ex version 3.00, a geochronological toolkit for Microsoft Excel. *Berkeley Geochronology Center Special Publ. 4*, May 30, (2003).
75. A.H. Jaffey, K.F. Flynn, L.E. Glendenin, W.C. Bentley, A.M. Essling, Precision measurements of half-lives and specific activities of $^{235}$U and $^{238}$U. *Physical Review* **C4**, 1889-1906 (1971).
76. C. Paton, J. Hellstrom, P. Paul, J. Woodhead, J. Hergt, Iolite: Freeware for the visualisation and processing of mass spectrometric data. *J Anal At Spectrom* **26**, 2508-2518 (2001).
77. D.J. Condon, N. McLean, S.R. Noble, S.A. Bowring, Isotopic composition ($^{238}$U/$^{235}$U) of some commonly used uranium reference materials. *Geochim Cosmochim Acta* **74**, 7127-7143 (2010).
78. F. Moynier, M. Le Borgne, High Precision Zinc Isotopic Measurements Applied to Mouse Organs. *Journal of Visualized Experiments* **99**, e52479 (2015).
79. H. Chen, P.S. Savage, F.-Z. Teng, R.T. Helz, F. Moynier, Zinc isotope fractionation during magmatic differentiation and the isotopic composition of the bulk Earth. *Earth Planet Sci Lett* **369-370**, 34-42 (2013).
80. E.R.D. Scott, A.N., Krot, in *Treatise on Geochemistry Vol. 1*, A.M. Davis, Ed. (Elsevier, 2014), pp. 65-137.

81. G.R. Huss, A.E. Rubin, J.N. Grossman, In *Meteorites and the Early Solar System II*, D. Lauretta, L. A. Leshin, H. Y. McSween, Eds. (Univ. Arizona Press, Tucson, 2006) pp. 567-586.
82. C. Göpel, G.Manhès, C.J. Allègre, U–Pb systematics of phosphates from equilibrated ordinary chondrites. *Earth Planet Sci Lett* **121**, 153-171 (1994).
83. A. Bouvier, J. Blichert-Toft, F. Moynier, J.D. Vervoort, F. Albarède, Pb–Pb dating constraints on the accretion and cooling history of chondrites. *Geochim Cosmochim Acta* **71**, 1583-1604 (2007).
84. E. Quirico, P.I. Raynal, M. Bourot-Denise, Pb–Pb dating constraints on the accretion and cooling history of chondrites. *Meteorit Planet Sci* **38**, 795-811 (2003).
85. L. Bonal, E. Quirico, M. Bourot-Denise, G. Montagnac, Pb-Pb dating constraints on the accretion and cooling history of chondrites. Determination of the petrologic type of CV3 chondrites by Raman spectroscopy of included organic matter. *Geochim Cosmochim Acta* **70**, 1849-1863 (2006).
86. M. Tatsumoto, R.J. Knight, C.J. Allègre, Time differences in the formation of meteorites as determined from the ratio of lead-207 to lead-206. *Science* **180**, 1279-1283. (1973).
87. J.S. Stacey, J.D. Kramers, Approximation of terrestrial lead isotope evolution by a two-stage model. *Earth Planet Sci Lett* **26**, 207-221 (1975).
88. E.R. Rambaldi, J.T. Wasson, Metal and associated phases in Bishunpur, a highly unequilibrated ordinary chondrite. *Geochim Cosmochim Acta* **45**, 1001-1015 (1981).
89. J.J. Bellucci, A.A. Nemchin, M.J. Whitehouse, J.F. Snape, P.A. Bland, G.K. Benedix, The Pb isotopic evolution of the Martian mantle constrained by initial Pb in Martian meteorites. *J Geophys Res (Planets)* **120**, 2224-2240 (2015).
90. M.J. Whitehouse, B.S. Kamber, C.M. Fedo, A. Lepland, Integrated Pb- and S-isotope investigation of sulphide minerals from the early Archaean of southwest Greenland. *Chem Geol* **222**, 112-131 (2005).
91. J.D. Woodhead, J.M. Hergt, Pb-Isotope Analyses of USGS Reference Materials Greenland. *Geostand Geoanal Res* **24**, 33-38 (2000).
92. Y. Amelin, A.N. Krot, I.D. Hutcheon, A.A. Ulyanov, Lead isotopic ages of chondrules and calcium-aluminium rich inclusions. *Science* **297**, 1678-1683 (2002).
93. S. Weyer, A.D. Anbar, A. Gerdes, G.W. Gordon, T.J. Algeo, E.A. Boyle, Natural fractionation of $^{238}$U/$^{235}$U. *Geochim Cosmochim Acta* **72**, 345-359 (2008).
94. C.E. Jilly-Rehak, G.R. Huss, K. Nagashima, Oxygen isotopes in secondary minerals in CR chondrites: Comparing components of different petrologic type. *43$^{rd}$ Lunar and Planetary Science Conference*, abstract #1662 (2015).
95. Y. Amelin, A.N. Krot, Pb isotopic age of the Allende chondrules. *Meteorit Planet Sci* **42**, 1321-1335 (2007).
96. J.N. Connelly, Y. Amelin, Y., A.N. Krot, M. Bizzarro, Chronology of the solar system's oldest solids. *Astrophys J Lett* **675**, L121-L124 (2008).

**Acknowledgments**

**General**: We thank Sasha Krot and Lars Buchhave for discussion on various aspects of this paper. All data needed to evaluate the conclusions in the paper are present in the paper and/or the Supplementary Materials. Additional data related to this paper may be requested from the authors. This is publication # 515 of the Nordsim facility.

**Funding:** Funded by grants from the Danish National Research Foundation (# DNRF97) and from the European Research Council (ERC Consolidator grant agreement 616027-Stardust2Asteroids) to M.B.

**Author contributions:** M.B. and J.N.C. designed research. J.B., J.N.C., M.J.W., E.A.P., L.B., F.M. and M.B. performed research. J.B., J.N.C., M.J.W., E.A.P., L.B., J.K.J., A.N., F.M. and M.B. analyzed data. M.B., J.B. and J.N.C. wrote the paper.

**Competing interests:** The authors declare that they have no competing interests.

# Figures and Tables

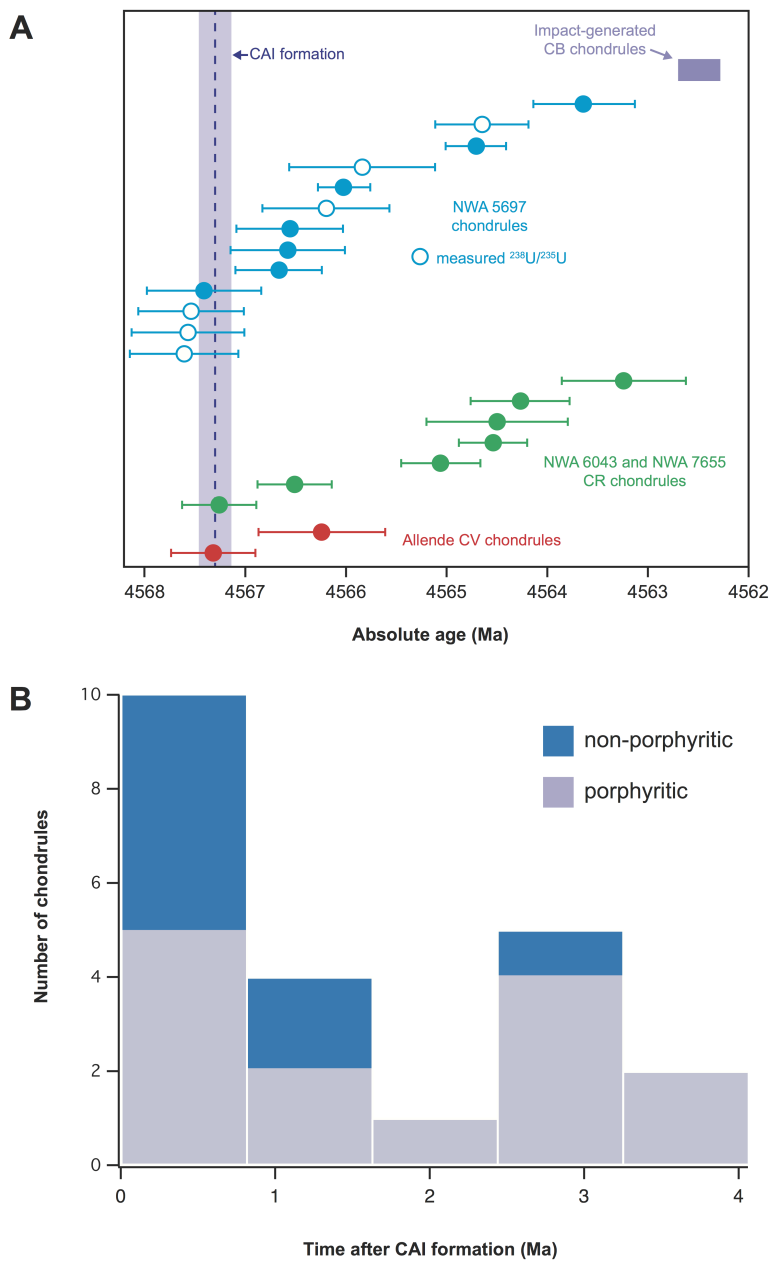

**Fig. 1. Absolute chronology of chondrule formation.** (**A**) Pb-Pb dates for individual chondrules from NWA 5697 (L3.10), NWA 6043 (CR2), NWA 7655 (CR2) and Allende (CV3). The Allende chondrules and three chondrules from NWA5697 were previously reported (6). The timing of CAI formation is accepted to be 4567.30±0.16 Ma (*6*). The CB chondrules are interpreted as having formed from colliding planetesimals at 4562.49±0.21 Ma (*31*). (**B**) Histogram depicting the the Pb-Pb age distribution of individual chondrules (N=22) relative to CAI formation. A full description of the methods used to collect the data reported in the paper are available in the Supplementary Materials.

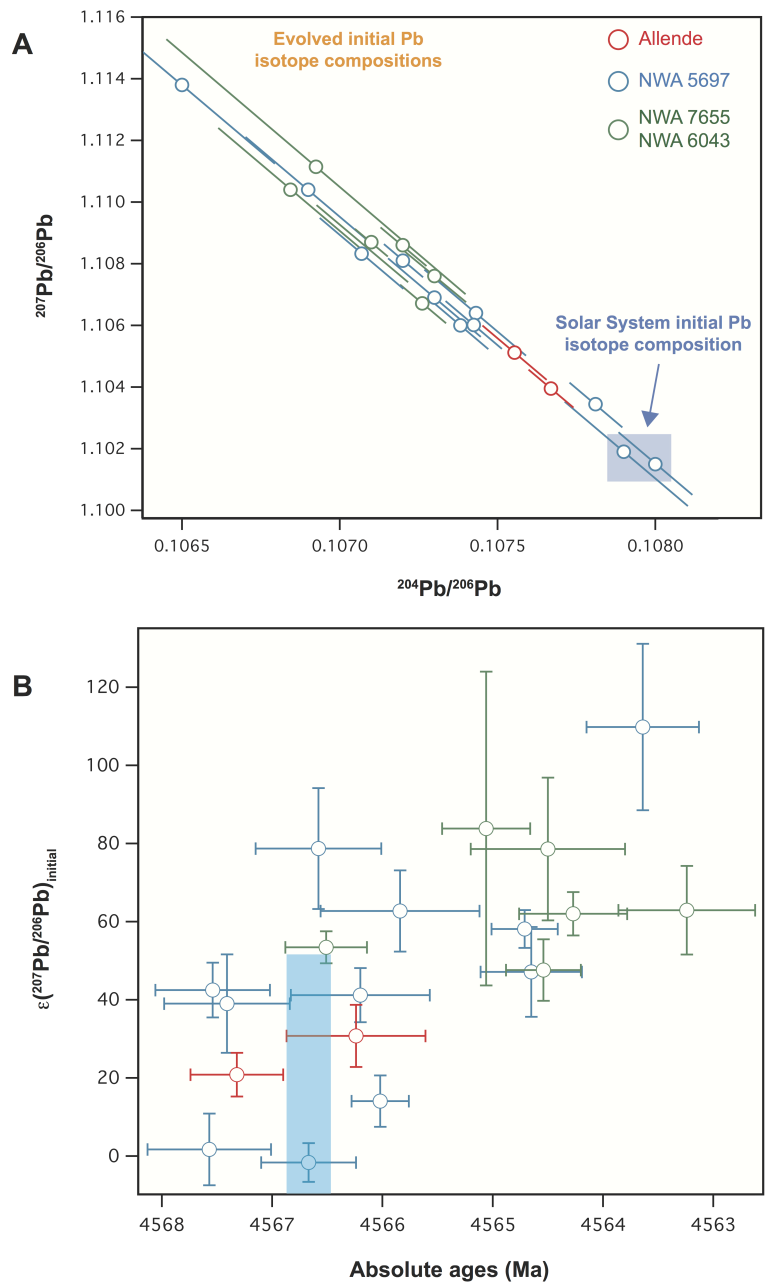

**Fig. 2. Lead isotope evolution diagrams.** (**A**) Initial Pb isotopic compositions of individual chondrules. The initial Pb isotope compositions are defined by the intersection of the individual isochrons and a Pb evolution array anchored on the solar system initial Pb isotope composition defined by chondrules 2-C1 and C1, which record the most primitive initial Pb isotope compositions. See Supplementary Materials for calculation of uncertainties on initial Pb isotope compositions. Individual chondrule data points have been displaced to the left and right hand side of the solar system initial Pb array for clarity. (**B**) Initial $^{207}$Pb/$^{206}$Pb compositions and age variation diagram. The ($^{207}$Pb/$^{206}$Pb)$_{initial}$ values are reported in per 10,000 deviations (ε-unit) from the composition defined by chondrules 2-C1 and C1. The blue box reflects the range of ε($^{207}$Pb/$^{206}$Pb)$_{initial}$ compositions of the >1 Myr chondrules back-calculated at 4566.8 Ma. The robustness of the ε($^{207}$Pb/$^{206}$Pb)$_{initial}$ and age correlation, including the effect of point selection of the individual regressions, has been statistically evaluated using a Monte Carlo approach (see Supplementary Materials).

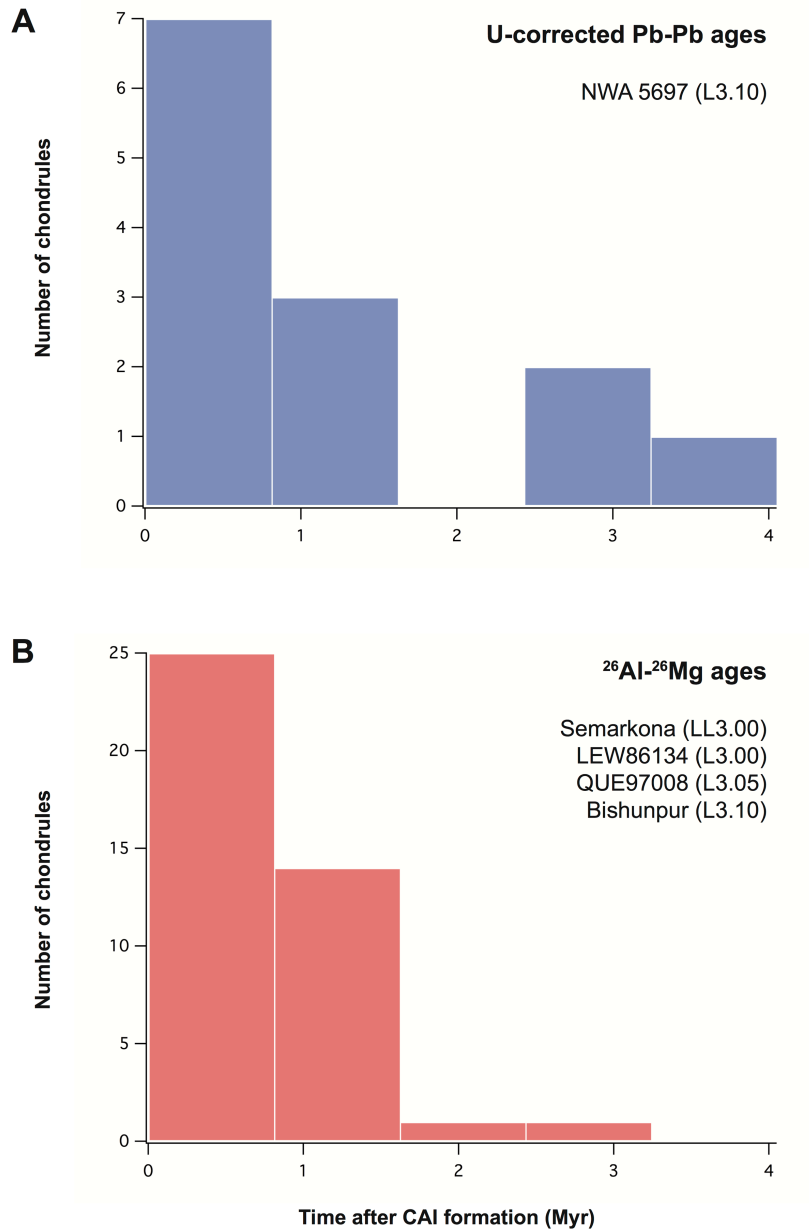

**Figure 3:** Histograms depicting the absolute and relative timing ages of chondrules from unequilibrated ordinary chondrites of low petrologic type (≤3.1) based on internal isochron relationships. (**A**) Pb-Pb dates of chondrules from the NWA 5697 ordinary chondrite (this study). (**B**) $^{26}$Al-$^{26}$Mg ages of chondrules from Semarkona, LEW 86134, QUE97008 and Bishunpur ordinary chondrites (*65-68*). The relative $^{26}$Al-$^{26}$Mg ages are calculated assuming that the precursor material from which these chondrules formed had a reduced initial $^{26}$Al/$^{27}$Al value corresponding to ~1.5×10$^{-5}$ (*33*). Three chondrules record initial $^{26}$Al/$^{27}$Al slightly higher than 1.5×10$^{-5}$ but are within analytical uncertainty of this estimate and, hence, have been assigned a T=0 formation age for simplicity.

Table 1. Summary of Pb-Pb dates, petrology, µ values, as well as $^{238}U/^{235}U$ and Zn isotope compositions of individual chondrules. Pb isotope data for the Allende chondrules and three chondrules from NWA 5697 (C1, C2 and C3) were previously reported by Connelly *et al.* (*6*) and are included here for completeness. P, porphyritic, NP, non-porphyritic, I, type I, II, type II. The $^{238}U/^{235}U$ uncertainties are propagated in the final age uncertainties. The Pb-Pb isochrons for three chondrules (5-C1, 5-C4 and 1-C2) project back to a modern terrestrial composition and, thus, accurate µ values cannot be calculated for these objects. The zinc isotope compositions are reported in δ notation, which reflects the permil (‰) deviations of the $^{66}Zn/^{64}Zn$ of the sample from the JMC Lyon standard.

| Chondrule | Texture and type | Pb-Pb age (Ma) | µ value | $^{238}U/^{235}U$ | δ$^{66}$Zn (‰) |
|---|---|---|---|---|---|
| *NWA 5697* | | | | | |
| 5-C1 | NP, II | 4567.61±0.54 | | 137.807±0.033 | |
| 2-C1 | NP, I | 4567.57±0.56 | 33 | 137.779±0.022 | −0.38±0.05 |
| 5-C2 | NP, II | 4567.54±0.52 | 21 | 137.756±0.029 | −1.07±0.05 |
| 5-C10 | NP, II | 4567.41±0.57 | 38 | 137.786±0.013 | −1.41±0.05 |
| C1 | NP, I | 4566.67±0.43 | 23 | 137.786±0.013 | |
| D-C3 | P, II | 4566.58±0.57 | 51 | 137.786±0.013 | −1.12±0.05 |
| 5-C4 | P, II | 4566.56±0.53 | | 137.786±0.013 | −1.15±0.05 |
| 3-C5 | P, II | 4566.20±0.63 | 28 | 137.807±0.026 | |
| C3 | NP, II | 4566.02±0.26 | 183 | 137.786±0.013 | |
| 11-C1 | NP, II | 4565.84±0.72 | 32 | 137.779±0.030 | |
| C2 | P, I | 4564.71±0.30 | 63 | 137.786±0.013 | |
| 11-C2 | NP, II | 4564.65±0.46 | 108 | 137.755±0.025 | −2.20±0.05 |
| 3-C2 | P, II | 4563.64±0.51 | 94 | 137.786±0.013 | −1.13±0.05 |
| *NWA 6043* | | | | | |
| 1-C2 | P, I | 4567.26±0.37 | | 137.786±0.013 | |
| 2-C2 | P, II | 4565.06±0.40 | 104 | 137.786±0.013 | |
| 2-C4 | P, II | 4564.50±0.70 | 93 | 137.786±0.013 | |
| 0-C1 | P, II | 4563.24±0.62 | 58 | 137.786±0.013 | |
| *NWA 7655* | | | | | |
| 1-C7 | P, II | 4566.51±0.37 | 2 | 137.786±0.013 | |
| 1-C2 | P, I | 4564.54±0.34 | 51 | 137.786±0.013 | |
| 1-C6 | P, II | 4564.27±0.49 | 9 | 137.786±0.013 | |
| *Allende* | | | | | |
| C30 | P, II | 4567.32±0.42 | 29 | 137.786±0.013 | |
| C20 | P, II | 4566.24±0.63 | 26 | 137.786±0.013 | |